\documentclass[%
 aip,
 jap,
 amsmath,amssymb,
 reprint,%
]{revtex4-1}

\usepackage{graphicx}
\usepackage{dcolumn}
\usepackage{bm}

\usepackage[utf8]{inputenc}
\usepackage[T1]{fontenc}
\usepackage{mathptmx}

\begin{document}

\preprint{AIP/123-QED}

\title{Differential coloration efficiency of electrochromic amorphous tungsten oxide as a function of intercalation level: Comparison between theory and experiment}

\author{Edgar A. Rojas-Gonz{\'a}lez}
 \email{edgar.rojas@angstrom.uu.se.}

\author{Gunnar A. Niklasson}%
\affiliation{ 
Department of Materials Science and Engineering, The {\AA}ngstr{\"o}m  Laboratory, Uppsala University, P.O. Box 534, SE-751 21 Uppsala, Sweden
}%

\date{\today}

\begin{abstract}
Optical absorption in amorphous tungsten oxide ($\textit{a}\mathrm{WO}_{3}$), for photon energies below that of the band gap, can be rationalized in terms of electronic transitions between localized states. For the study of this phenomenon, we employed the differential coloration efficiency concept, defined as the derivative of the optical density with respect to the inserted charge. We also made use of its extension to a complex quantity in the context of frequency-resolved studies. Combined \textit{in situ} electrochemical and optical experiments were performed on electrochromic $\textit{a}\mathrm{WO}_{3}$ thin films for a wide lithium intercalation range using an optical wavelength of $810~\mathrm{nm}$ ($1.53~\mathrm{eV}$). Quasi-equilibrium measurements were made by chronopotentiometry (CP). Dynamic frequency-dependent measurements were carried out by simultaneous electrochemical and color impedance spectroscopy (SECIS). The differential coloration efficiency obtained from CP changes sign at a critical intercalation level. Its response exhibits an excellent agreement with a theoretical model that considers electronic transitions between $\mathrm{W}^{4+}$, $\mathrm{W}^{5+}$, and $\mathrm{W}^{6+}$ sites. For the SECIS experiment, the low-frequency limit of the differential coloration efficiency shows a general trend similar to that from CP. However, it does not change sign at a critical ion insertion level. This discrepancy could be due to degradation effects occurring in the films at high $\mathrm{Li}^+$ insertion levels. The methodology and results presented in this work can be of great interest both for the study of optical absorption in disordered materials and for applications in electrochromism.
\end{abstract}

\maketitle


\section{\label{sec1:level1}Introduction}
Electrochromic (EC) materials modify their optical properties upon an electrical stimulus. They can be employed, for example, in energy-efficient applications like smart windows.\cite{Granqvist2014}
The EC material is usually prepared in a thin film form and arranged between an electronic back contact and a reservoir of small ions, typically a proton- or lithium-containing electrolyte. Reversible intercalation of ions into the EC host can be achieved by the application of an external voltage or current. The optical modulation is caused by the electrons that enter the EC film from the back contact due to charge neutrality requirements associated with the intercalation.\cite{Granqvist1995}
Tungsten oxide ($\mathrm{WO}_{3}$) is one of the most studied and used EC materials.\cite{Granqvist_2016,Granqvist2019} In normal operational conditions, it colors under ion insertion, which is denoted as cathodic coloration. The opposite behavior is called anodic coloration and can be observed, for instance, in nickel oxide.\cite{Granqvist1995}

Amorphous tungsten oxide ($\textit{a}\mathrm{WO}_{3}$) exhibits an optical absorption peak centered at higher energies with respect to that of crystalline $\mathrm{WO}_{3}${\textemdash}around $1.2$ to $1.3~\mathrm{eV}$ for the former, and $0.6$ to $0.7~\mathrm{eV}$ for the latter.\cite{Niklasson2007}
The localization of electrons is enhanced by the presence of disorder in the amorphous case, and this can explain the blue shift of the absorption peak with respect to the crystalline counterpart.\cite{Bryksin1982}

Optical absorption of disordered systems like $\textit{a}\mathrm{WO}_{3}$ (for photon energies below that of the band gap) can be rationalized in terms of either intervalence transfer\cite{Hush1967} or small-polaron absorption processes,\cite{Austin1969,Schirmer1974,Bottger1985} and both models rely on electronic transitions between localized states. In this context, the site-saturation theory (SS) developed by Denesuk and Uhlmann\cite{Denesuk1996} models the optical density dependence on the intercalation level $x${\textemdash}which, for $\textit{a}\mathrm{WO}_{3}$ and a lithium-containing electrolyte is given by the $\mathrm{Li}/\mathrm{W}$ ratio. It considers electronic transitions between neighboring $\mathrm{W}^{5+}$ and $\mathrm{W}^{6+}$ sites and only allows for a maximum occupancy level of one electron per tungsten atom. An extended site-saturation theory (ES){\textemdash}that admits up to two electrons per tungsten atom by including $\mathrm{W}^{4+}$ sites{\textemdash}was introduced by Berggren, Jonsson, and Niklasson.\cite{Berggren2007} Both SS and ES derive the number of available electronic transitions from probability considerations for a system in equilibrium.

The coloration efficiency is an important parameter for EC materials and devices.\cite{Granqvist1995} It is defined as the change in optical density per unit inserted charge. It can be regularly found in literature either as the initial slope in a plot of the optical density against the inserted charge, or in terms of parameters describing a bleached and a colored state{\textemdash}see for example Ref. [\onlinecite{Park2019}], and Ref. [\onlinecite{Sorar2019}], respectively. The present work differs from the previous approaches because it makes use of the differential coloration efficiency, defined as the derivative of the optical density with respect to the inserted charge. This definition has been used elsewhere\cite{Denesuk1996,Denesuk1997,Denesuk1997a,Denesuk2000} in relation to $\textit{a}\mathrm{WO}_{3}$. However, to our knowledge, it has not been employed at high intercalation levels ($x\gtrsim0.5$), where interesting phenomena can be observed in the optical absorption.\cite{Berggren2007} Additionally, the differential coloration efficiency has been extended to a complex quantity in the context of frequency-resolved studies.\cite{SusanaInesCordobaTorresi1990,Kim1997,Kim1997a,Bueno2008,Agrisuelas2009,Agrisuelas2009a,Agrisuelas2015}

In this work, we present \textit{in situ} optical measurements of $\textit{a}\mathrm{WO}_{3}$ thin films for a wide intercalation range using an optical wavelength of $810~\mathrm{nm}$ ($1.53~\mathrm{eV}$){\textemdash}which is close to the maximum of the optical absorption peak of $\textit{a}\mathrm{WO}_{3}$. A special emphasis is given to the differential coloration efficiency, that was obtained both from a quasi-equilibrium method{\textemdash}low-current chronopotentiometry\cite{Mattsson1998} (CP){\textemdash}and a dynamic technique that studies small oscillations around an equilibrium state{\textemdash}simultaneous electrochemical and color impedance spectroscopy\cite{Rojas-Gonzalez2019} (SECIS). We find intriguing differences between the results obtained by the two techniques. Moreover, the experimental results are compared to the site-saturation theories.

In the following section, we describe in detail the experimental procedures used in this work. Next, we introduce the theoretical concepts related to the differential coloration efficiency and its connection with the CP and SECIS measurements and the site-saturation theories. Subsequently, the experimental results are presented and discussed. Finally, we add some conclusions and remarks.

\section{\label{sec2:level1}Experimental setup and procedures}

\subsection{\label{sec2:level2_1}Electrode preparation}
Tungsten oxide ($\mathrm{WO}_{3}$) thin films were prepared by reactive DC magnetron sputtering using a deposition system based on a Balzers UTT 400 unit (with a distance from target to substrate of $13~\mathrm{cm}$). The target was a $5$-$\mathrm{cm}$-$\mathrm{diameter}$ $99.95$-$\%$-$\mathrm{pure}$ metallic $\mathrm{W}$ disc. The deposition system was initially evacuated to $\sim10^{-7}~\mathrm{Torr}$. Then, the target was pre-sputtered in $99.9997$-$\%$-$\mathrm{pure}$ $\mathrm{Ar}$ for about $5~\mathrm{min}$ to clean it from surface contamination. The deposition onto unheated substrates was performed at a constant discharge power of $240~\mathrm{W}$ in an atmosphere of $\mathrm{O}_2$ ($99.998$-$\%$-$\mathrm{pure}$) and $\mathrm{Ar}$ with total pressure of $30~\mathrm{mTorr}${\textemdash}with $\mathrm{Ar}$, and $\mathrm{O}_2$ flows of $50$, and $22~\mathrm{ml}/\mathrm{min}$, respectively. The substrates were rotated during deposition to assure film thickness homogeneity. During the same deposition run, glass pre-coated with conducting $\mathrm{In}_2\mathrm{O}_3$:$\mathrm{Sn}$ (ITO; $15~\Omega/\mathrm{sq}$), and glassy carbon plates were used as substrates for opto-electrochemical measurements, and Rutherford backscattering spectrometry (RBS) analysis, respectively. The RBS measurements were performed at the Tandem Laboratory at Uppsala University using $2~\mathrm{MeV}$ $^4\mathrm{He}$ ions backscattered at an angle of $170^\circ$. The density and elemental contents of the films were determined by fitting the RBS spectra to a model for the film-substrate system using the SIMNRA program.\cite{Mayer1999} The $\mathrm{WO}_{3}$ thin films presented an amorphous structure, as confirmed by X-ray diffraction (XRD) patterns that were collected by a Siemens D5000 diffractometer using Cu $K\alpha$ radiation.
The thicknesses of the $\textit{a}\mathrm{WO}_{3}$ films were determined by stylus profilometry using a Bruker DektakXT instrument. The samples discussed below were prepared in two different deposition runs{\textemdash}namely, batch-1 and batch-2. For batch-1 (batch-2), 
the RBS analysis, together with thickness determinations, provided a density $\rho$ of $4.81\pm0.01~\mathrm{g}~\mathrm{cm}^{-3}$ ($4.73\pm0.01~\mathrm{g}~\mathrm{cm}^{-3}$), a tungsten number density $N_\mathrm{W}$ of $[1.242\pm0.002]\times10^{22}~\mathrm{cm}^{-3}$ ($[1.218\pm0.002]\times10^{22}~\mathrm{cm}^{-3}$), and an $\mathrm{O}/\mathrm{W}$ ratio of $3.11\pm0.03$ ($3.13\pm0.03$).

\subsection{\label{sec2:level2_2}Electrochemical and optical setup}
The electrochemical measurements were performed in an argon-filled glove box ($\mathrm{H}_2\mathrm{O}$ level $< 0.6~\mathrm{ppm}$) using a standard three-electrode setup in a quartz cuvette. The electrolyte was $1~\mathrm{M}~\mathrm{LiClO}_4$ dissolved in propylene carbonate. Both the reference and counter electrodes consisted of lithium foils, and a fresh $\textit{a}\mathrm{WO}_{3}$ sample acted as the working electrode (WE) for each experiment. The \textit{in situ} optical transmittance of the WE was measured using an optical setup described in Ref [\onlinecite{Rojas-Gonzalez2019}]. Here, a collimated beam of light was collected by a photodetector after passing through the two parallel faces of the quartz cuvette, the electrolyte, and the WE. The light source was a LED with peak wavelength at 810 nm (M810F2, Thorlabs). 

For the CP experiment, we used a BioLogic SP-200 potentiostat. In this case, the output voltage of the photodetector was feed into an analog input of the SP-200 potentiostat. The setup for SECIS measurements used in this work was described in detail elsewhere,\cite{Rojas-Gonzalez2019} including the estimation of the uncertainties of the optical signal. For the SECIS method, a sinusoidal excitation voltage is applied to the WE (at a stationary equilibrium state of interest) and the resulting sinusoidal current and transmittance responses are measured simultaneously. The frequency-dependent electrical measurements were carried out by an electrochemical interface (SI-1286, Solartron) connected to a frequency response analyzer (FRA; SI-1260, Solartron). The transmittance was measured simultaneously, as described above, and the output from the photodetector was fed into the current input of the FRA. The frequency-dependent relations between the amplitudes and relative phases of the sinusoidal excitation and the responses provide physical insights about the electrical and optical properties of the electrochromic system under study.

\subsection{\label{sec2:level2_3}CP experiment}
Here, we describe the experimental sequence followed during the quasi-equilibrium CP experiment{\textemdash}using a sample from batch-1 with film thickness $d=294\pm6~\mathrm{nm}$ and active area of the electrode $A=1.26\pm0.08~\mathrm{cm}^2$. First, the initial open circuit potential (OCP) of the fresh WE was measured, and its value was $3.72~\mathrm{V}~\mathrm{vs.}~\mathrm{Li}/\mathrm{Li}^+$. Following the definition given later in Eq.~(\ref{sec4:eq:6}), the photodetector output voltage measured at this initial bleached state of the $\textit{a}\mathrm{WO}_{3}$ WE was defined as the $100$-$\%$-level of transmittance. Next, cyclic voltammetry (CV) during three cycles in the voltage range of $2.0$-$4.0~\mathrm{V}~\mathrm{vs.}~\mathrm{Li}/\mathrm{Li}^+$ was performed with the first linear sweep toward the low potential limit. This technique started and finished at the value of the initial OCP. For all the fresh samples studied in this work, the first CV cycle was consistently different from the second and third ones, that were almost identical. The CV measurement was immediately followed by a chronopotentiometry measurement, in which a constant current of $1~\mathrm{\mu A}${\textemdash}in the direction of $\mathrm{Li}^+$ intercalation into the WE{\textemdash}was applied for a time span of $72~\mathrm{h}$. For the resulting current density value ($\lesssim1~\mathrm{\mu A}~\mathrm{cm}^{-2}$), the equilibrium bias potential can be approximated by the one measured during the CP technique.\cite{Mattsson1998} The time-dependent intercalation level $x(t)$ was calculated as follows

\begin{equation}
x(t)=(zedAN_\mathrm{W})^{-1}\int_0^t \mathrm{d}t' I(t'),\label{sec2:eq:1}
\end{equation}

with $z$ the valence of the intercalated ion (for a monovalent cation like $\mathrm{Li}^+$ we have $z=1$), $e$ the elementary charge, and $I(t)$ the current during the chronopotentiometry measurement (which, for convenience, was set to begin at $t=0$). The time dependence of the intercalation level (related to the CP experiment) is dropped in the remainder of this work.

\subsection{\label{sec2:level2_4}SECIS experiment}

The results from two SECIS experiments{\textemdash}with different transition protocols{\textemdash}are discussed in the current work. The first type of measurement (denoted SECIS-1) was performed on a WE from batch-1 (with $d=294\pm4~\mathrm{nm}$ and $A=1.28\pm0.09~\mathrm{cm}^2$), and its experimental sequence is described as follows. First, the OCP was measured{\textemdash}giving a value of $3.82~\mathrm{V}~\mathrm{vs.}~\mathrm{Li}/\mathrm{Li}^+$ for the fresh WE. Then, a CV during three cycles was performed in the voltage range of $2.0$-$4.0~\mathrm{V}~\mathrm{vs.}~\mathrm{Li}/\mathrm{Li}^+$ starting and finishing at the OCP obtained just before the CV{\textemdash}the first potential sweep went toward the low potential edge. Next, a linear potential sweep was applied down to the desired bias potential value, followed by a potentiostatic polarization treatment for $20~\mathrm{min}${\textemdash}letting the WE reach its electrochemical steady-state condition. Thereafter, the SECIS measurements were carried out at the intended bias potential in the frequency range between $10~\mathrm{mHz}$ and $30~\mathrm{kHz}$ with an excitation voltage amplitude of $20~\mathrm{mV}$ root-mean-square (rms). For each frequency, the integration was performed during $4$ cycles after a delay of $1$ cycle. Thereafter, the same process was repeated for the next bias potential value. The impedance measurements were made (in descending order) for bias potential values of $3.15$, $2.90$, $2.60$, $2.10$, and $1.50~\mathrm{V}~\mathrm{vs.}~\mathrm{Li}/\mathrm{Li}^+$.

For the second type of measurement (denoted SECIS-2), we used a transition protocol{\textemdash}between the relevant bias potential values{\textemdash}that resembled the conditions of the CP experiment. Here, a sample from batch-2 (with $d=299\pm2~\mathrm{nm}$ and $A=1.09\pm0.08~\mathrm{cm}^2$) was used. The studied bias potentials, measured in descending order, were $3.15$, $2.90$, $2.60$, $2.37$, $2.15$, $1.95$, $1.83$, $1.73$, and $1.64~\mathrm{V}~\mathrm{vs.}~\mathrm{Li}/\mathrm{Li}^+$. Here, the two first sequences were identical to those of SECIS-1{\textemdash}namely, the measurement of the initial OCP of the fresh WE (which, in this case, was $3.65~\mathrm{V}~\mathrm{vs.}~\mathrm{Li}/\mathrm{Li}^+$) and the CV. Then, the WE was driven to the bias potential value of interest by a CP technique using a current of $1~\mathrm{\mu A}$ in the direction of lithium intercalation. Subsequently, a SECIS measurement was directly performed as described for SECIS-1. Successively, the CP and SECIS techniques were repeated for the remaining bias potential values. Additionally, the OCP of the WE was measured for $10~\mathrm{h}$ after the end of the last SECIS measurement (which was performed at the bias potential of $1.64~\mathrm{V}~\mathrm{vs.}~\mathrm{Li}/\mathrm{Li}^+$). 

\section{\label{se3:level1}Theory}

Here, we define the concept of differential coloration efficiency in general and in terms of the parameters relevant to SECIS and the site-saturation theories. It is implied that the quasi-equilibrium CP case directly follows the general considerations.

\subsection{\label{sec3:level2_1}Differential coloration efficiency: General concepts}

The optical density $OD(\lambda,x)$ at the optical wavelength $\lambda$ and intercalation level $x$ is defined as $OD(\lambda,x)=d\alpha(\lambda,x)${\textemdash}with $\alpha(\lambda,x)$ the absorption coefficient. It can be related to the total transmittance $T_t(\lambda,x)$ and reflectance $R_t(\lambda,x)$ by the expression\cite{Hong1989}

\begin{equation}
OD(\lambda,x)=\mathrm{ln}\left(\frac{1-R_t(\lambda,x)}{T_t(\lambda,x)}\right).\label{sec4:eq:1}
\end{equation}

The differential coloration efficiency $K(\lambda,x)$ is given by

\begin{equation}
K(\lambda,x)=\mathrm{d}OD(\lambda,x)/\mathrm{d}q,\label{sec4:eq:2}
\end{equation}

with the inserted charge per unit area

\begin{equation}
q=edN_\mathrm{W}x.\label{sec4:eq:3}
\end{equation}

Plugging Eq.~(\ref{sec4:eq:1}) into Eq.~(\ref{sec4:eq:2}), we obtain

\begin{equation}
K(\lambda,x)=-\frac{1}{T_t(\lambda,x)}\frac{\mathrm{d}T_t(\lambda,x)}{\mathrm{d}q}-\frac{1}{[1-R_t(\lambda,x)]}\frac{\mathrm{d}R_t(\lambda,x)}{\mathrm{d}q},\label{sec4:eq:4}
\end{equation}

with the right-hand side depending on the total transmittance, reflectance, and their derivatives with respect to $q$. 

Next, we will simplify Eq.~(\ref{sec4:eq:4}) according to the specific experimental conditions used in this work. Optical measurements on samples similar to those used here\cite{Berggren2007,Triana2015} show that the amplitude of the interference fringes in the reflectance{\textemdash}around $810~\mathrm{nm}${\textemdash}decreases significantly when the intercalation level increases from $x=0$ to some value in between $0.04$ and $0.15$, and a relatively constant reflectance is observed for $x\gtrsim0.15$ (at least up to $x=1.8$, which is the highest intercalation level presented in Ref. [\onlinecite{Berggren2007}]). In addition, it is worth mentioning that electrodeposition of Li-containing species on the $\textit{a}\mathrm{WO}_{3}$ films{\textemdash}that can abruptly modify the optical properties{\textemdash}is only relevant at bias potentials below about $1.1~\mathrm{V}~\mathrm{vs.}~\mathrm{Li}/\mathrm{Li}^+$,\cite{Isidorsson2000} which lie outside the range studied in the present work. 

Hence, for $T_t(\lambda,x)$ and $[1-R_t(\lambda,x)]$ of the same order of magnitude and an almost constant $R_t(\lambda,x)$, the last term of the right-hand side of Eq.~(\ref{sec4:eq:4}) can be neglected. We thus get

\begin{equation}
K(\lambda,x)=-\frac{1}{T_t(\lambda,x)}\frac{\mathrm{d}T_t(\lambda,x)}{\mathrm{d}q}.\label{sec4:eq:5}
\end{equation}

The form of Eq.~(\ref{sec4:eq:5}) makes $K(\lambda,x)$ independent of the normalization factor of the transmittance. Then, we can replace $T_t(\lambda,x)$ by a modified transmittance $T(\lambda,x)$ defined as

\begin{equation}
T(\lambda,x) = I_\mathrm{L}(\lambda,x)/I_\mathrm{L}(\lambda,x=0),\label{sec4:eq:6}
\end{equation}

with $I_\mathrm{L}(\lambda,x)$ the intensity of the transmitted light. In this case, the intensity measured at the bleached state of $\textit{a}\mathrm{WO}_{3}$, $I_\mathrm{L}(\lambda,x=0)$, is set as the $100$-$\%$-level. This definition is convenient because it directly shows the changes upon intercalation with respect to the $x=0$ reference case. Hereinafter, the quantity defined in Eq.~(\ref{sec4:eq:6}) will be referred to simply as transmittance.

\begin{figure*}
\includegraphics[scale=0.50]{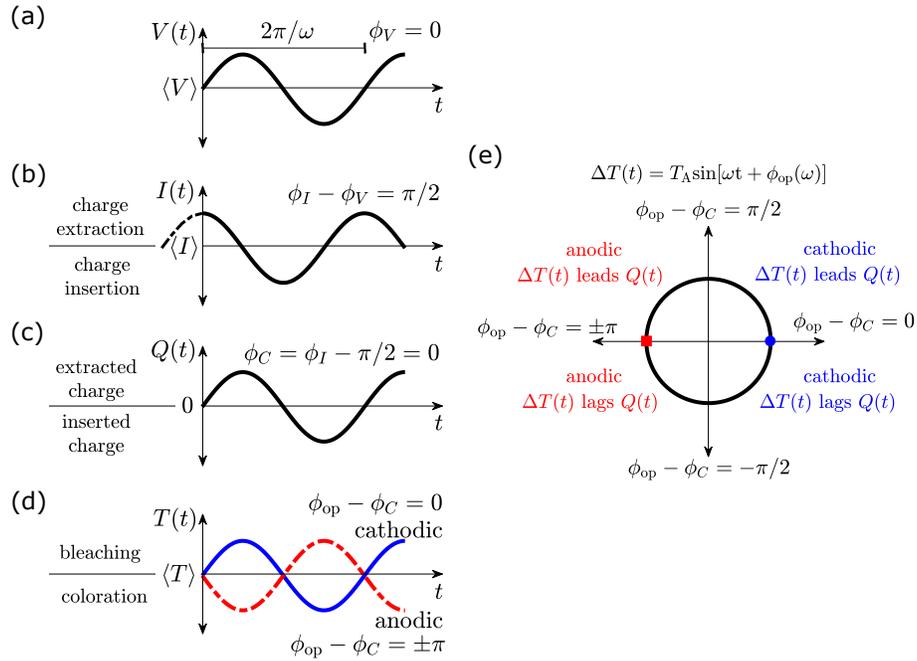}
\caption{\label{fig:1} Schematic representation of the sinusoidal excitation and responses for an illustrative case corresponding to a pure capacitive behavior ($\phi_I(\omega)-\phi_V=\pi/2$) with perfectly synchronized coloration and charge. (a) Sinusoidal excitation voltage. (b) Sinusoidal current response{\textemdash}the regions (with respect to $\langle I \rangle$) with charge extraction and insertion are indicated in the plot. (c) Sinusoidal charge response{\textemdash}with respect to the equilibrium condition, the vertical half planes corresponding to net extracted and inserted charge are portrayed. (d) Sinusoidal transmittance response with $\phi_\mathrm{op}(\omega)-\phi_C(\omega)=0$, and $\phi_\mathrm{op}(\omega)-\phi_C(\omega)=\pm \pi$ for perfectly synchronized cathodic, and anodic behavior, respectively{\textemdash}the vertical regions with bleaching and coloration (with respect to $\langle T \rangle$) are depicted in the plot. (e) Relative phase $\phi_\mathrm{op}(\omega)-\phi_C(\omega)$ represented as points on the unit circle{\textemdash}the relative phases for the perfectly synchronized cathodic, and anodic cases shown in (d) are indicated by a blue circle, and a red square, respectively. In (e), the main characteristics of each quadrant are indicated{\textemdash}namely, the cathodic or anodic behavior and the relation between $\Delta T(t)=T_\mathrm{A}(\omega) \mathrm{sin}[\omega t+ \phi_\mathrm{op}(\omega)]$ and $Q(t)$. These characteristics also depict the behavior of the complex frequency-dependent differential coloration efficiency $\tilde{K}(\omega)$ defined in Eq.~(\ref{sec4:eq:13}) because its phase $\phi_K(\omega)$ is equal to $\phi_\mathrm{op}(\omega)-\phi_C(\omega)$.}
\end{figure*}

\subsection{\label{sec3:level2_2}Differential coloration efficiency: SECIS}

Following the same definitions presented in Ref. [\onlinecite{Rojas-Gonzalez2019}], the oscillatory excitation voltage $V(t)$ applied during the SECIS experiments at the stationary equilibrium bias potential $\langle V \rangle$ and circular frequency $\omega=2\pi f$ (with $f$ the corresponding linear frequency) is expressed as

\begin{equation}
V(t) = \langle V \rangle+V_\mathrm{A} \mathrm{sin}(\omega t+ \phi_V),\label{sec4:eq:7}
\end{equation}

with $V_\mathrm{A}$, and $\phi_V$ the amplitude, and phase of the sinusoidal excitation, respectively. For convenience, we set $\phi_V=0$. Assuming that linearity holds, the resulting current $I(t)$, and transmittance $T(t)$ responses are respectively given by

\begin{eqnarray}
I(t) &=& \langle I \rangle+I_\mathrm{A}(\omega) \mathrm{sin}[\omega t+ \phi_I(\omega)],\label{sec4:eq:8}\\
T(t) &=& \langle T \rangle+T_\mathrm{A}(\omega) \mathrm{sin}[\omega t+ \phi_\mathrm{op}(\omega)],\label{sec4:eq:9}
\end{eqnarray}

with $I_\mathrm{A}(\omega)$ and $T_\mathrm{A}(\omega)$, and $\phi_I(\omega)$ and $\phi_\mathrm{op}(\omega)$ the frequency-dependent amplitudes, and phases of the corresponding sinusoidal responses, respectively. The quantities $\langle I \rangle$, and $\langle T \rangle$ are the stationary equilibrium values for the current, and transmittance, respectively. Note that the current $I(t)$ is defined positive when it flows from the WE toward the electrolyte{\textemdash}that is, the direction of $\mathrm{Li}^+$ extraction from the film{\textemdash}and negative in the opposite case{\textemdash}that is, the direction of $\mathrm{Li}^+$ insertion into the film. The oscillating charge response $Q(t)${\textemdash}due to the sinusoidal excitation voltage{\textemdash}can be obtained from Eq.~(\ref{sec4:eq:8}) by

\begin{eqnarray}
Q(t) &=& \int_0^t \mathrm{d}t'[I(t')- \langle I \rangle]\nonumber\\
&=& Q_\mathrm{A}(\omega)\mathrm{sin}[\omega t+ \phi_C(\omega)],\label{sec4:eq:10}
\end{eqnarray}

with $ Q_\mathrm{A}(\omega)= I_\mathrm{A}(\omega)/\omega$ and $\phi_C(\omega)=\phi_I(\omega)-\pi/2$. The sign convention for the current $I(t)$ implies that negative, and positive values of $Q(t)$ correspond respectively to a net inserted, and extracted charge with respect to the stationary equilibrium condition given by $\langle V \rangle$, $\langle I \rangle$, and $\langle T \rangle${\textemdash}note that this is opposite to $q$ from Eq.~(\ref{sec4:eq:3}), which increases with inserted charge. The definitions and conventions mentioned above are better visualized by looking at the representative illustrative case depicted in Fig.~\ref{fig:1}. It corresponds to a sinusoidal excitation voltage at circular frequency $\omega$, see Fig.~\ref{fig:1}(a), with the current exhibiting a pure capacitive response with $\phi_I(\omega)-\phi_V=\pi/2$, see Fig.~\ref{fig:1}(b). Correspondingly, the charge depicted in Fig.~\ref{fig:1}(c) presents $\phi_C(\omega)=\phi_I(\omega)-\pi/2=0$, being in phase with the voltage. Figure~\ref{fig:1}(d) portrays a transmittance that is perfectly synchronized with the charge{\textemdash}here, the corresponding cathodic (with $\phi_\mathrm{op}(\omega)-\phi_C(\omega)=0$) and anodic (with $\phi_\mathrm{op}(\omega)-\phi_C(\omega)=\pm \pi$) coloration cases are depicted.

Returning to the general case, using the amplitudes and phases from Eqs. (\ref{sec4:eq:9}) and (\ref{sec4:eq:10}), we can define the complex frequency-dependent transmittance $\tilde{T}(\omega)$, and charge $\tilde{Q}(\omega)$ respectively as

\begin{eqnarray}
\tilde{T}(\omega)&=&T_\mathrm{A}(\omega)e^{i \phi_{\mathrm{op}}{(\omega)}}.\label{sec4:eq:11}\\
\tilde{Q}(\omega)&=&Q_\mathrm{A}(\omega)e^{i \phi_C{(\omega)}},\label{sec4:eq:12}
\end{eqnarray}

The complex frequency-dependent differential coloration efficiency $\tilde{K}(\omega)$ related to SECIS (at optical wavelength $\lambda$ and intercalation level $x$ specified for each case) is given by 

\begin{equation}
\tilde{K}(\omega) = A\langle T \rangle^{-1}[\tilde{T}(\omega)/\tilde{Q}(\omega)]=|\tilde{K}(\omega)|e^{i\phi_K(\omega)},\label{sec4:eq:13}
\end{equation}

with $\tilde{Q}(\omega)/A$ the complex frequency-dependent charge per unit area, $|\tilde{K}(\omega)|=A\langle T \rangle^{-1}[T_\mathrm{A}(\omega)/Q_\mathrm{A}(\omega)]$, and $\phi_K(\omega)=\phi_{\mathrm{op}}(\omega)-\phi_{C}(\omega)$. Figure~\ref{fig:1}(e) presents a schematic description of the main characteristics of the complex coloration efficiency according to the value of the relative phase $\phi_\mathrm{op}(\omega)-\phi_C(\omega)${\textemdash}namely, the cathodic or anodic nature and whether the sinusoidal transmittance response leads or lags the sinusoidal charge response. 

The quantity $\tilde{K}(\omega)$ provides information about the dynamics of the intercalation process. In particular, it follows from the conventions described above that positive, and negative values of the real part of $\tilde{K}(\omega)$ correspond to cathodic (coloration upon ion insertion), and anodic (coloration upon ion extraction) behavior, respectively (Fig.~\ref{fig:1}(e)). In addition, for the specific cases of $\phi_{K}(\omega)=0$ ($\tilde{K}(\omega)$ real and positive), and $\phi_{K}(\omega)=\pm \pi$ ($\tilde{K}(\omega)$ real and negative) the sinusoidal transmittance response is perfectly correlated, and anticorrelated with the sinusoidal charge, respectively (Fig.~\ref{fig:1}(d)). Moreover, for $0<\phi_K(\omega)<\pi$ (upper half of the Nyquist plot with $\mathrm{Im}\{\tilde{K}(\omega)\}>0$), and $-\pi<\phi_K(\omega)<0$ (lower half of the Nyquist plot with $\mathrm{Im}\{\tilde{K}(\omega)\}<0$) the sinusoidal transmittance response leads, and lags the sinusoidal charge, respectively (Fig.~\ref{fig:1}(e)). Furthermore, for $\omega \rightarrow 0$ we assume that the sinusoidal transmittance and charge responses are synchronized and that the frequency-dependent method resembles the quasi-equilibrium case{\textemdash}provided that both methods are probing the same reversible phenomena and that the linearity condition is satisfied in the frequency-dependent case. Then, the low-frequency intersection of $\tilde{K}(\omega)$ with the real axis{\textemdash}in a Nyquist plot{\textemdash}is expected to be equal to $K(\lambda,x)$ from Eq.~(\ref{sec4:eq:5}). 

\subsection{\label{sec3:level2_3}Differential coloration efficiency: Site-saturation theories}

The optical density $OD(\lambda,x)$ can be expressed as

\begin{equation}
OD(\lambda,x)=\Sigma_{j\in B}OD_j(\lambda,x),\label{sec4:eq:14}
\end{equation}

where the summation runs over the elements of the set of absorbing processes (denoted by $B$), and $OD_j(\lambda,x)$ is the optical density of element $j$.

The SS theory deals with only one absorbing process{\textemdash}that is, $\mathrm{W}^{5+}\rightarrow \mathrm{W}^{6+}$ transitions.\cite{Denesuk1996} Here, the optical density is given by

\begin{equation}
OD^\mathrm{SS}(\lambda,x)=A^\mathrm{SS}(\lambda)P^\mathrm{SS}(x),\label{sec4:eq:15}
\end{equation}

with $A^\mathrm{SS}(\lambda)$ a wavelength-dependent coefficient that characterizes the absorption strength, and $P^\mathrm{SS}(x)=(x-x^2)$ the probability that a transition takes place at a random tungsten site{\textemdash}this model is mathematically defined up to $x=1$. 

The ES theory includes three possible absorption processes{\textemdash}those are, the electronic transitions $\mathrm{W}^{5+}\rightarrow\mathrm{W}^{6+}$, $\mathrm{W}^{4+}\rightarrow\mathrm{W}^{5+}$, and $\mathrm{W}^{4+}\rightarrow\mathrm{W}^{6+}$. In the case of stoichiometric $\mathrm{WO}_3$, they present probabilities of transition at a random tungsten site given by\cite{Berggren2007}

\begin{subequations}
\label{sec4:eq:16}
\begin{equation}
P_{56}^\mathrm{ES}(x)=x(1-x/2)^3,\label{sec4:eq:16a}
\end{equation}
\begin{equation}
P_{45}^\mathrm{ES}(x)=(x^3/4)(1-x/2),\label{sec4:eq:16b}
\end{equation}
\begin{equation}
P_{46}^\mathrm{ES}(x)=(x^2/4)(1-x/2)^2,\label{sec4:eq:16c}
\end{equation}
\end{subequations}

respectively. This model is mathematically defined up to $x=2$. Here, the optical density can be expressed as

\begin{equation}
OD^\mathrm{ES}(\lambda,x)=A^\mathrm{ES}_{56}(\lambda)P^\mathrm{ES}_{56}(x)+A^\mathrm{ES}_{45}(\lambda)P^\mathrm{ES}_{45}(x)+A^\mathrm{ES}_{46}(\lambda)P^\mathrm{ES}_{46}(x),\label{sec4:eq:17}
\end{equation}

with the contribution of each process being weighted by their corresponding wavelength-dependent absorption strength coefficient.

The theoretical differential coloration efficiency according to SS $K^\mathrm{SS}(\lambda,x)$ is obtained by plugging Eq.~(\ref{sec4:eq:15}) into Eq.~(\ref{sec4:eq:2}) and using the relation given in Eq.~(\ref{sec4:eq:3}). It reads

\begin{equation}
K^\mathrm{SS}(\lambda,x)=C^\mathrm{SS}(\lambda)~\mathrm{d}P^\mathrm{SS}(x)/\mathrm{d}x,\label{sec4:eq:18}
\end{equation}

with $C^\mathrm{SS}(\lambda)\equiv (edN_\mathrm{W})^{-1}A^\mathrm{SS}(\lambda)$ and

\begin{equation}
\mathrm{d}P^\mathrm{SS}(x)/\mathrm{d}x = 1-2x.\label{sec4:eq:19}
\end{equation}

Similarly, plugging Eq.~(\ref{sec4:eq:17}) into Eq.~(\ref{sec4:eq:2}), the theoretical differential coloration efficiency according to ES $K^\mathrm{ES}(\lambda,x)$ can be expressed as

\begin{eqnarray}
K^\mathrm{ES}(\lambda,x)&=&C^\mathrm{ES}_{56}(\lambda)~\mathrm{d}P^\mathrm{ES}_{56}(x)/\mathrm{d}x+C^\mathrm{ES}_{45}(\lambda)~\mathrm{d}P^\mathrm{ES}_{45}(x)/\mathrm{d}x \nonumber\\
& &+C^\mathrm{ES}_{46}(\lambda)~\mathrm{d}P^\mathrm{ES}_{46}(x)/\mathrm{d}x,\label{sec4:eq:20}
\end{eqnarray}

with

\begin{subequations}
\label{sec4:eq:21}
\begin{equation}
\mathrm{d}P^\mathrm{ES}_{56}(x)/\mathrm{d}x=(1/2)(x-2)^2(1/2-x),\label{sec4:eq:21a}
\end{equation}
\begin{equation}
\mathrm{d}P^\mathrm{ES}_{45}(x)/\mathrm{d}x=(1/2)x^2(3/2-x),\label{sec4:eq:21b}
\end{equation}
\begin{equation}
\mathrm{d}P^\mathrm{ES}_{46}(x)/\mathrm{d}x=(1/4)x(x-1)(x-2).\label{sec4:eq:21c}
\end{equation}
\end{subequations}

\begin{figure}[b]
\includegraphics[scale=0.50]{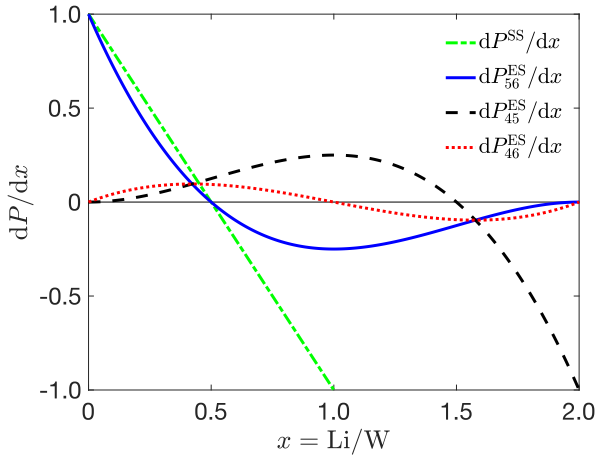}
\caption{\label{fig:2} Derivative of the probability of transition at a random tungsten site as a function of intercalation level for the processes considered by the SS and ES theories.}
\end{figure}

The dependence on the intercalation level of the derivative of the probabilities of transition{\textemdash}given by Eqs. (\ref{sec4:eq:19}) and (\ref{sec4:eq:21}){\textemdash}is depicted in Fig.~\ref{fig:2}. $\mathrm{d}P^\mathrm{SS}(x)/\mathrm{d}x$ consists of a straight line with slope $-2$ that crosses the horizontal axis at $x=0.5$. The term $\mathrm{d}P^\mathrm{ES}_{46}(x)/\mathrm{d}x$ is equal to zero at $x=0$, $x=1$, and $x=2${\textemdash}it is positive for $0<x<1$ and negative for $1<x<2$. Similar to the SS curve, $\mathrm{d}P^\mathrm{ES}_{56}(x)/\mathrm{d}x$ is positive for $0<x<0.5$ and vanishes at $x=0.5$; however, it differs significantly from the SS case for $x>0.5$, presenting a minimum at $x=1$ and vanishing again at $x=2$. Finally, $\mathrm{d}P^\mathrm{ES}_{45}(x)/\mathrm{d}x$ corresponds to the same curve as $\mathrm{d}P^\mathrm{ES}_{56}(x)/\mathrm{d}x$ but mirrored both around the horizontal axis and $x=1$.

\begin{figure*}
\includegraphics[scale=0.42]{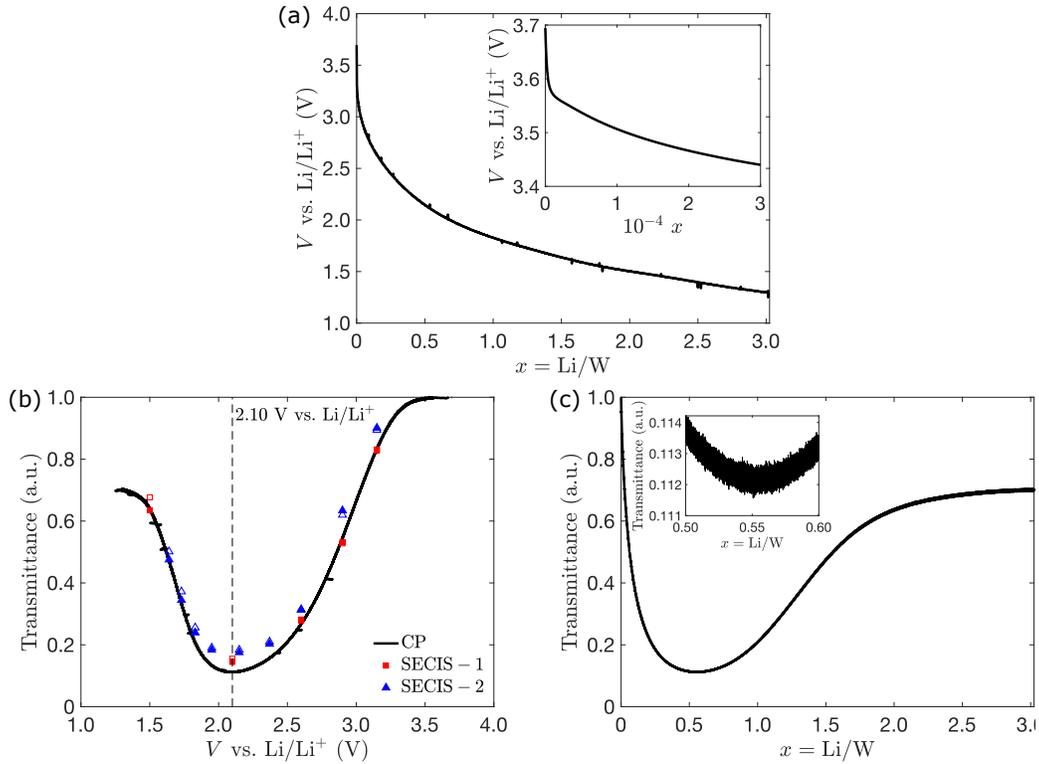}
\caption{\label{fig:3} (a) Bias potential during the CP experiment as a function of intercalation level{\textemdash}the inset displays the data on an expanded scale for $0 \leq x \leq 3\times 10^{-4}$. The transmittance during the CP experiment is portrayed in (b), and (c) as a function of bias potential, and intercalation level, respectively. The dashed line in (b) depicts the bias potential around which the transmittance minimum is located (that is, $2.10~\mathrm{V}~\mathrm{vs.}~\mathrm{Li}/\mathrm{Li}^+$). The stationary equilibrium transmittance values during the SECIS measurements for the SECIS-1 (red squares) and SECIS-2 (blue triangles) experiments are shown in (b){\textemdash}the filled, and unfilled symbols correspond to the stationary equilibrium transmittance values at the beginning, and end of each SECIS measurement, respectively. The inset in (c) displays an expanded scale of the region around the transmittance minimum located at about $x=0.56$.}
\end{figure*} 

\section{\label{se4:level1}Results and discussion}
The bias potential measured during the CP experiment decreases with the intercalation level for the whole experimental range studied in this work, see Fig.~\ref{fig:3}(a). It presents an abrupt drop at low values of $x$ and, in general, tends to diminish its rate of change with increasing intercalation level. The corresponding transmittance is shown in Fig.~\ref{fig:3}(b) as a function of bias potential. It exhibits maxima at low and high potentials and between them a conspicuous minimum located at about $2.10~\mathrm{V}~\mathrm{vs.}~\mathrm{Li}/\mathrm{Li}^+$. Considering Fig.~\ref{fig:3}(c), it should be noted that low, and high intercalation levels correspond to high, and low bias potentials, respectively. Hence, the transmittance as a function of intercalation level follows a similar trend as that observed in Fig.~\ref{fig:3}(b). It exhibits its maximum at $x=0$ and decreases monotonically with $x$ up to about $x=0.56$. Then, it increases up to about $x=2$ before flattening for higher intercalation levels. Note that the stationary equilibrium transmittance during the SECIS measurements, depicted in Fig.~\ref{fig:3}(b), presents the same behavior as the corresponding CP curve. Also, notice that the stationary equilibrium transmittance is relatively stable during the SECIS measurements that were performed at $\langle V \rangle > 2.10~\mathrm{V}~\mathrm{vs.}~\mathrm{Li}/\mathrm{Li}^+${\textemdash}see the vertical displacement between the unfilled and filled symbols in Fig.~\ref{fig:3}(b). On the other hand, a positive transmittance drift is observed at $\langle V \rangle \leq 2.10~\mathrm{V}~\mathrm{vs.}~\mathrm{Li}/\mathrm{Li}^+$ and this effect is more pronounced at the lowest bias potential values.               

\begin{figure}
\includegraphics[scale=0.43]{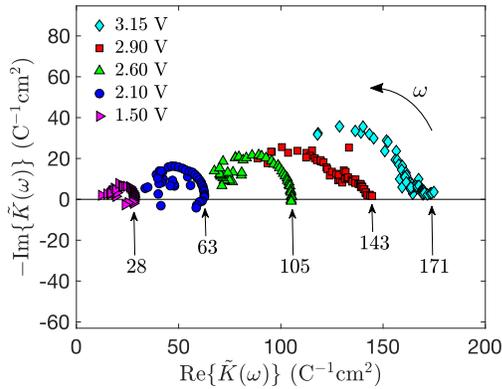}
\caption{\label{fig:4} Nyquist plot of the complex differential coloration efficiency obtained in the SECIS-1 experiment for bias potential values{\textemdash}with respect to $\mathrm{Li}/\mathrm{Li}^+${\textemdash}of $3.15~\mathrm{V}$ (cyan diamonds), 2.90 V (red squares), $2.60~\mathrm{V}$ (green up-facing-triangles), $2.10~\mathrm{V}$ (blue circles), and $1.50~\mathrm{V}$ (magenta right-facing triangles). The low-frequency intersections with the real axis are indicated in units of $\mathrm{C}^{-1}\mathrm{cm}^2$. For each spectrum, the high-frequency region with signal-to-noise ratio of the optical signal smaller than $5$ is not displayed. In all cases, the highest frequency presented in the plot was $\sim 10~\mathrm{Hz}$ and the lowest one was $10~\mathrm{mHz}$. The curved arrow indicates the direction of increasing frequency.}
\end{figure}

\begin{figure*}
\includegraphics[scale=0.50]{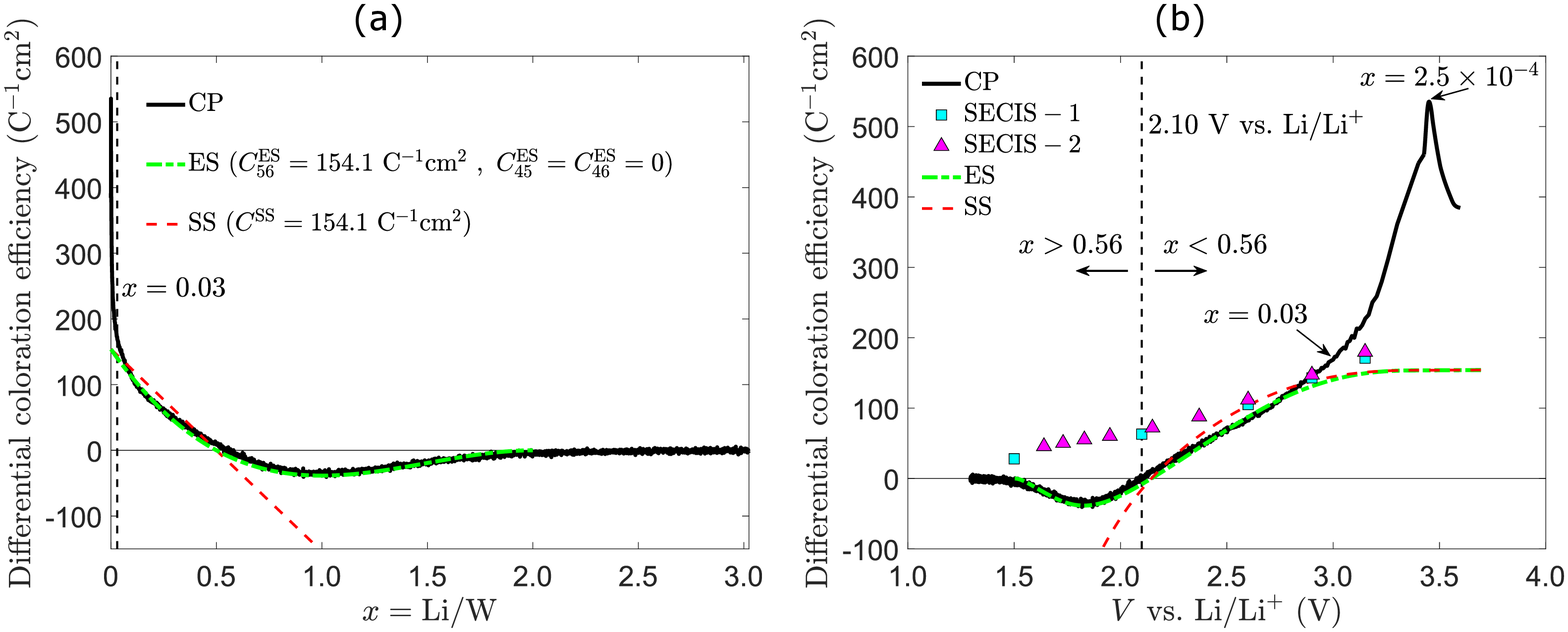}
\caption{\label{fig:5}Differential coloration efficiency as a function of intercalation level (a) and bias potential (b). In both plots, the experimental result from CP (black solid curve) was compared to the theoretical prediction from the SS (red dashed curve) and ES (green dash-dotted curve) theories using the fitting coefficients given in (a). The vertical dashed line in (a) depicts the position of the $x=0.03$ intercalation level. Panel (b) includes the low-frequency intersection of $\tilde{K}(\omega)$ with the real axis for each impedance measurement of the SECIS-1 (cyan squares) and SECIS-2 (magenta up-facing-triangles) experiments. The vertical dashed line in (b) depicts the bias potential around which the differential coloration efficiency obtained from CP is equal to zero (that is, $2.10~\mathrm{V}~\mathrm{vs.}~\mathrm{Li}/\mathrm{Li}^+$){\textemdash}$x$ increases and decreases, with respect to the vertical dashed line (corresponding to $x \approx 0.56$), toward the respective directions shown by the arrows. Some representative $x$ values are indicated in (b). The experimental standard deviations of the SECIS data are of the order of $0.5~\mathrm{C}^{-1}\mathrm{cm}^{2}$ (or lower), much smaller than the size of the symbols.}
\end{figure*}

The Nyquist plot of the complex differential coloration efficiency measured during the SECIS-1 experiment is depicted in Fig.~\ref{fig:4}. Here, each spectrum shows an arc-shaped dispersion, starting closer to the origin at high frequencies and converging around the real axis at low frequencies. The lower the bias potential the closer the low-frequency intersection with the real axis is to the origin. In general, for all the bias potentials shown here, the data present a positive real part of $\tilde{K}(\omega)$ and a negative imaginary part of $\tilde{K}(\omega)$. According to Eq.~(\ref{sec4:eq:13}), this corresponds to a phase $\phi_K(\omega)=\phi_\mathrm{op}(\omega)-\phi_C(\omega)$ located in the fourth quadrant of Fig.~\ref{fig:1}(e){\textemdash}that is, in terms of the response signals, a cathodic coloration with a sinusoidal transmittance that lags the sinusoidal charge. The data points with positive imaginary part of $\tilde{K}(\omega)$ show a minor effect at the lowest frequencies at low bias potentials, which was nevertheless significant. The spectra from the SECIS-2 experiment, not shown here, exhibit the same qualitative behavior as those from SECIS-1.        

The comparison between the experimental differential coloration efficiency{\textemdash}from CP and SECIS{\textemdash}and the theoretical site-saturation models is depicted in Fig.~\ref{fig:5}. For each SECIS measurement, the plotted value corresponds to the low-frequency intersection of $\tilde{K}(\omega)$ with the real axis. In Fig.~\ref{fig:5}(a), the CP curve presents a maximum close to $x=0$. From this point, it decreases monotonously with the intercalation level and vanishes at about $x=0.56$. Then, it becomes negative and keeps decreasing up to about $x=1$, where it reaches a minimum. Thereafter, it starts increasing up to about $x=2$ and flattens around zero for higher intercalation levels. The initial behavior of the CP curve up to about $x=0.03$ could not be reconciled with the site-saturation theories. In view of this, the ES theory was fitted to the CP experimental data in the range $0.03<x<2$, giving a coefficient $C_{56}^\mathrm{ES}=154.1\pm1.1~\mathrm{C}^{-1}\mathrm{cm}^2${\textemdash}which, according to Eqs.~(\ref{sec4:eq:20}) and (\ref{sec4:eq:21}), also corresponds to the value of the differential coloration efficiency at $x=0$. In the fit reported in Fig.~\ref{fig:5}(a), we put $C_{45}^\mathrm{ES}=C_{46}^\mathrm{ES}=0$ because only a marginal improvement was achieved when these parameters were released. This is motivated by the fact that, for an optical wavelength of $810~\mathrm{nm}$ ($1.53~\mathrm{eV}$), the optical absorption is mainly dominated by $\mathrm{W}^{5+} \rightarrow \mathrm{W}^{6+}$ transitions.\cite{Berggren2007} The SS model has as a fitting parameter the value of the differential coloration efficiency at $x=0$. Due to the discrepancy between the site-saturation theories and the CP data at low intercalation levels, the same value found for $C_{56}^\mathrm{ES}$ was assigned to $C^\mathrm{SS}$.     

After the previous considerations, and focusing on the range with $0.03<x<2$, we can now compare the experimental and theoretical results presented in Fig.~\ref{fig:5}(a). The most remarkable observation is that the ES model exhibits an excellent agreement with the experimental results. Additionally, the SS follows the general trend of the CP curve up to about $x=0.5$, but diverges from it for higher intercalation levels. Regarding the fitting shown in Fig.~\ref{fig:5}(a) for the ES, the contributions to the optical absorption from transitions involving the $\mathrm{W}^{4+}$ sites were ignored{\textemdash}that is, their respective coefficients were fixed to zero. Actually, the inclusion of $C_{45}^\mathrm{ES}$ and $C_{46}^\mathrm{ES}$ in the fitting gives rise to a small shift (from $x=0.5$) of the point at which the ES curve changes sign and also a slight deviation from zero at $x=2$. It is also important to remark that the effect of the $\mathrm{W}^{4+}$ sites is implicitly included in the $\mathrm{d}P_{56}^\mathrm{ES}(x)/\mathrm{d}x$ term. In addition, only the ES model could correctly follow the CP curve for a wide intercalation range. As a result, the previous arguments clearly suggest that the consideration of the $\mathrm{W}^\mathrm{4+}$ sites{\textemdash}in the theoretical models{\textemdash}is necessary to appropriately reproduce the experimental response. 

Figure~\ref{fig:5}(b) presents the same information shown in Fig.~\ref{fig:5}(a) and the SECIS data as a function of bias potential. Here, the CP curve follows the same trend as that explained above. Particularly, it vanishes at about  $V=2.10~\mathrm{V}~\mathrm{vs.}~\mathrm{Li}/\mathrm{Li}^+$ ($x=0.56$), exhibits a minimum at about $V=1.83~\mathrm{V}~\mathrm{vs.}~\mathrm{Li}/\mathrm{Li}^+$ ($x=1$), and flattens around zero for $V \lesssim 1.50~\mathrm{V}~\mathrm{vs.}~\mathrm{Li}/\mathrm{Li}^+$ ($x \gtrsim 2$). On top of that, the bias potential representation allows to distinguish a peak located at about $V=3.45~\mathrm{V}~\mathrm{vs.}~\mathrm{Li}/\mathrm{Li}^+$ ($x=2.5 \times 10^{-4}$). The high bias potential region above $3.00~\mathrm{V}~\mathrm{vs.}~\mathrm{Li}/\mathrm{Li}^+$ (low intercalation levels with $x<0.03$) presents a significant discrepancy between the CP curve and the site-saturation theories. This region corresponds to an abrupt drop of the bias potential with the intercalation level{\textemdash}which can be observed in Fig.~\ref{fig:3}(a) for low values of $x$. In connection to this, it is important to remember that the electron energy increases in the direction of decreasing bias potential. A previous study of $\textit{a}\mathrm{WO}_{3}$, employing the same electrolyte and similar samples to those used in the present work, approximated the position of the conduction band edge at about $3.05~\mathrm{V}~\mathrm{vs.}~\mathrm{Li}/\mathrm{Li}^+$.\cite{Niklasson2010} Therefore, the features at potentials above this value in Fig.~\ref{fig:5}(b) occur at electron energies below the conduction band edge. Hence, the abrupt drop of the bias potential at low intercalation levels can be assigned to effects due to localized electronic states in the $\textit{a}\mathrm{WO}_{3}$ band gap{\textemdash}that is, a band tail that extends into the band gap and possibly other defect states superimposed to it. As a result, we argue that the response of the experimental differential coloration efficiency from CP at the high bias potential range is dominated by these states{\textemdash}whose optical absorption evidently presents different characteristics than those described by the site-saturation theories used in this work. Furthermore, this argument suggests that the site saturation theory should only be applied at potentials below {$3.05~\mathrm{V}~\mathrm{vs.}~\mathrm{Li}/\mathrm{Li}^+$}, where it indeed is observed to provide a good fit.        

\begin{figure}[t]
\includegraphics[scale=0.45]{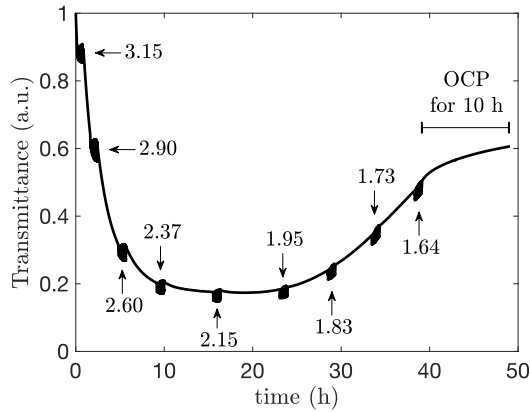}
\caption{\label{fig:6} Transmittance during the SECIS-2 experiment (after the CV sequence) as a function of time. The features corresponding to the SECIS measurements are indicated by arrows labeled by their respective stationary equilibrium bias potential in units of $\mathrm{V}$ with respect to $\mathrm{Li}/\mathrm{Li}^+$. The $10~\mathrm{h}$ time span during which the OCP was monitored (after the last SECIS measurement) is depicted in the plot.}
\end{figure}

Concerning the SECIS data depicted in Fig.~\ref{fig:5}(b), a direct comparison to the CP data can be done by using the intercalation level as a reference. Thus, we assigned to each bias potential the intercalation level given by the one-to-one correspondence between $V$ and $x$ from the CP experiment depicted in Fig.~\ref{fig:3}(a). This choice is supported by the fact that the transmittance that was simultaneously measured during the SECIS sequences was consistent with that from the CP experiment, see Fig.~\ref{fig:3}(b). Also, we note the resemblance between the transmittance measured during the SECIS-2 experiment shown in Fig.~\ref{fig:6} and that depicted in Fig.~\ref{fig:3}(c){\textemdash}it is important to notice that the intercalation level in Fig.~\ref{fig:6} is proportional to the elapsed time between SECIS measurements. The previous observations tell us that the underlying quasi-equilibrium transmittance of the SECIS measurements{\textemdash}and its dependence on the intercalation level{\textemdash}is equivalent to that measured during the CP experiment. 

Now, we focus again on the information shown in Fig.~\ref{fig:5}(b). The SECIS-1 data are, in general, of the same order of magnitude as those from CP and follow a similar trend{\textemdash}namely, they decrease with decreasing bias potential. Subsequently, the SECIS-2 variant was used in order to more closely reproduce the conditions of the CP experiment. Notably, the new measurements confirmed the results obtained by SECIS-1, as depicted in Fig.~\ref{fig:5}(b). However, the change of sign of the differential coloration efficiency observed in the CP experiment was not reproduced. In order to check the veracity of the SECIS data presented in Fig.~\ref{fig:5}(b), additional time-resolved measurements (not shown here) were performed on an $\textit{a}\mathrm{WO}_{3}$ WE at bias potential values of $2.60$, $2.10$, and $1.50~\mathrm{V}~\mathrm{vs.}~\mathrm{Li}/\mathrm{Li}^+$, with a superimposed sinusoidal excitation voltage signal with amplitude of $20~\mathrm{mV}$ rms and frequency of $0.5~\mathrm{Hz}$. For all these cases, the time-dependent data clearly showed a sinusoidal transmittance response that was correlated with the sinusoidal charge response{\textemdash}that is, a cathodic coloration behavior. Therefore, the fact that the SECIS data does not change sign in Fig.~\ref{fig:5}(b) is related to the sample response and not to a measurement artifact.    

The difference in behavior between the quasi-equilibrium measurement and the low-frequency limit of the dynamic method is an interesting result that deserves further attention. Figure~\ref{fig:6} shows that the SECIS experiments at potentials below $2~\mathrm{V}~\mathrm{vs.}~\mathrm{Li}/\mathrm{Li}^+$ were not carried out under steady-state conditions. It is seen that the transmittance level was not stable during the SECIS measurements, but instead experienced a continuous increase. The transmittance continued to increase under open-circuit conditions for $10~\mathrm{h}$ after the measurements, as also seen in Fig.~\ref{fig:6}. It is well-known that the electrochromic properties of $\mathrm{WO}_{3}$ are not stable in the potential region below $2~\mathrm{V}~\mathrm{vs.}~\mathrm{Li}/\mathrm{Li}^+${\textemdash}that is, the ability to modulate the transmittance decreases with time during color-bleach cycling. This degradation is believed to be due to trapping of the inserted $\mathrm{Li}^+$ ions,\cite{Wen2015,Bisquert2002} or to the formation of transparent Li oxides and/or hydroxides,\cite{Bressers1998} and is probably responsible, at least in part, for the continuously increasing average transmittance during the SECIS measurements. Hence, the SECIS response is the outcome of at least two competing effects{\textemdash}namely, the electrochromic coloration/bleaching and an underlying degradation kinetics. In addition, effects of the electronic density of states may be important at high $x$. It should be noted that an underlying assumption in the site saturation theories is that the involved electron states do not have significantly different energies. Clearly, more studies are needed to elucidate the relation between the mechanisms responsible for the quasi-static and dynamic coloration in $\textit{a}\mathrm{WO}_{3}$.    

\section{\label{sec5:level1}Conclusions}
Combined electrochemical and optical measurements on $\textit{a}\mathrm{WO}_{3}$ were made in a wide intercalation range. The differential coloration efficiency concept gives relevant insights into the description of electrochromic materials. Also, it can be used to compare experimental results to theories that model the coloration in $\textit{a}\mathrm{WO}_{3}$. In the quasi-equilibrium regime, $\textit{a}\mathrm{WO}_{3}$ makes a transition from cathodic to anodic coloration at a critical intercalation level, and this can be reconciled with an optical absorption caused by electronic transitions between localized states. In fact, combined chronopotentiometry and optical measurements were found to be in excellent agreement with an extended site saturation theory. This site-saturation framework considers the effect of $\mathrm{W}^{4+}$, $\mathrm{W}^{5+}$, and $\mathrm{W}^{6+}$ sites on the optical absorption of $\textit{a}\mathrm{WO}_{3}$. Particularly, the consideration of $\mathrm{W}^{4+}$ sites was needed to reproduce the experimental results in a wide intercalation level range. A difference in the quasi-equilibrium and dynamic coloration efficiency of $\textit{a}\mathrm{WO}_{3}$ was observed, primarily at low bias potentials. This is probably due to an underlying degradation process that affects the stability of the conditions during the dynamical electrochemical and opto-impedance measurements. The methods and results of the present paper pave the way for both fundamental studies of the optical absorption of disordered electrochromic materials and the development of practical electrochromic applications.            

\begin{acknowledgments}
This work was supported by a grant from the Swedish Research Council (No. VR-2016-03713). E. A. Rojas-Gonz{\'a}lez is grateful for the support from the University of Costa Rica. Daniel Primetzhofer, Marcos Moro, and the staff at the Tandem Laboratory at Uppsala University are thanked for assistance with RBS measurements. Support by VR-RFI (contract 2017-00646-9) and the Swedish Foundation for Strategic Research (contract RIF14-0053) for accelerator operation is gratefully acknowledged.
\end{acknowledgments}


%

\end{document}